# Resilience management during large-scale epidemic outbreaks

Emanuele Massaro[1,2,3], Alexander Ganin[1,4], Nicola Perra[5,6,7], Igor Linkov[1*], Alessandro Vespignani[6,7,8*]

[1]U.S. Army Corps of Engineers – Engineer Research and Development Center, Environmental Laboratory, Concord, MA, 01742, USA , [2]Senseable City Laboratory, Massachusetts Institute of Technology, 77 Massachusetts Avenue, Cambridge, MA 02139, USA, [3]HERUS Lab, École Polytechinque Fédérale de Lausanne (EPFL), CH-1015 Lausanne, Switzerland, [4]University of Virginia, Department of Systems and Information Engineering, Charlottesville, VA, 22904, USA, [5]Business School of Greenwich University, London, UK, [6]Laboratory for the Modeling of Biological and Socio-Technical Systems, Northeastern University, Boston, MA 02115, USA, [7]Institute for Scientific Interchange, 10126 Torino, Italy, [8]Institute for Quantitative Social Sciences at Harvard University, Cambridge, MA 02138, USA

**\*Corresponding Authors**

IL, U.S. Army Corps of Engineers – Engineer Research and Development Center, Environmental Laboratory, Concord, MA, 01742, USA; email: Igor.Linkov@usace.army.mil

AV, Sternberg Distinguished University Professor, Department of Physics, College of Computer and Information Sciences, Bouvé College of Health Sciences, Northeastern University. 177 Huntington Ave, 10th Floor**,** Northeastern University · 360 Huntington Ave., Boston, Massachusetts 02115; email: a.vespignani@neu.edu

# Abstract

Assessing and managing the impact of large-scale epidemics considering only the individual risk and severity of the disease is exceedingly difficult and could be extremely expensive. Economic consequences, infrastructure and service disruption, as well as the recovery speed, are just a few of the many dimensions along which to quantify the effect of an epidemic on society's fabric. Here, we extend the concept of resilience to characterize epidemics in structured populations, by defining the system-wide critical functionality that combines an individual's risk of getting the disease (disease attack rate) and the disruption to the system's functionality (human mobility deterioration). By studying both conceptual and data-driven models, we show that the integrated consideration of individual risks and societal disruptions under resilience assessment framework provides an insightful picture of how an epidemic might impact society. In particular, containment interventions intended for a straightforward reduction of the risk may have net negative impact on the system by slowing down the recovery of basic societal functions. The presented study operationalizes the resilience framework, providing a more nuanced and comprehensive approach for optimizing containment schemes and mitigation policies in the case of epidemic outbreaks.

# Introduction

Data-driven models of infectious diseases [1–15] are increasingly used to provide real- or near-real-time situational awareness during disease outbreaks. Indeed, notwithstanding the limitations inherent to predictions in complex systems, mathematical and computational models have been used to forecast the size of epidemics [16–19], assess the risk of case importation across the world [10,14,20], and communicate the risk associated to uncurbed epidemics outbreaks [21–23]. Despite contrasting opinions on the use of modelling in epidemiology [24], in the last few years a large number of studies have employed them to evaluate disease containment and mitigation strategies as well as to inform contingency plans for pandemic preparedness [11,13,15,24,25]. Model-based epidemic scenarios in most cases focus on the ``how many and for how long?'' questions. Furthermore, mitigation and containment policies are currently evaluated in the modelling community by the reduction they produce on the attack rate (number of cases) in the population. These studies aim at identifying best epidemic management strategies but typically neglect the epidemic and mitigation impact on the societal functions overall.

The evaluation of vulnerabilities and consequences of epidemics is a highly dimensional complex problem that should consider societal issues such as infrastructures and services disruption, forgone output, inflated prices, crisis-induced fiscal deficits and poverty [26,27]. Therefore, it is important to broaden the model-based approach to epidemic analysis, expanding the purview by including measures able to assess the system resilience, i.e. response of the entire system to disturbances, their aftermath, the outcome of mitigation as well as the system's recovery and retention of functionality [28–30]. Most important, operationalizing resilience [29–31] must include the temporal dimension; i.e. a system's recovery and retention of functionality in the face of adverse events [30,32–35]. The assessment and management of system resilience to epidemics must, therefore, identify the critical functionalities of the system and evaluate the temporal profile of how they are maintained or recover in response to adverse events.

Even though the assessment and management of adverse events resilience of complex systems is the subject of active research [32,33,35,36], its integration in the computational analysis of epidemic threats is still largely unexplored [27,37,38].

Here, we introduce a resilience framework to the analysis of the global spreading of an infectious disease in structured populations. We simulate the spread of infectious diseases across connected populations, and monitor the system–level response to the epidemic by introducing a definition of engineering resilience that compounds both the disruption caused by the restricted travel and social distancing, and the incidence of the disease. We find that while intervention strategies, such as restricting travel and encouraging self-initiated social distancing, may reduce the risk to individuals of contracting the disease, they also progressively degrade population mobility and reduce the critical functionality thus making the system less resilient. Our numerical results show a transition point that signals an abrupt change of the overall resilience in response to these mitigation policies. Consequently, containment measures that reduce risk may drive the system into a region associated with long-lasting overall disruption and low resilience. Interestingly, this region is in proximity of the global invasion threshold of the system, and it is related to the slowing down of the epidemic progression. Our study highlights that multiple dimensions of a socio-technical system must be considered in epidemic management and sets forward a new framework of potential interest in analyzing contingency plans at the national and international levels.

## Results

We provide a general framework for the analysis of the system-level resilience to epidemics by initially considering a metapopulation network (Figure 1A). In this case we consider a system made of $V$ distinct subpopulations. These form a network in which each subpopulation $i$ is made of $N_i$ individuals and is connected to a set $k_i$ of other subpopulations. A complete description of the networked systems is given in the Methods section. The notation and the description of the parameters used in our simulations are reported in Table 1.

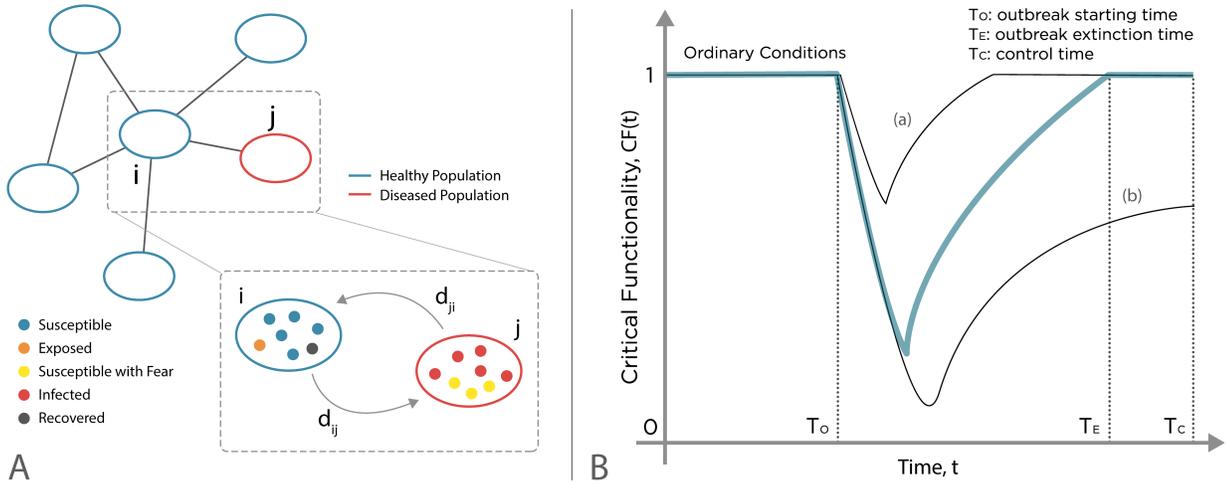

**Figure 1. Schematic representation of the metapopulation model.** The system is composed of a network of subpopulations or patches, connected by diffusion processes. Each patch contains a population of individuals who are characterized with respect to their stage of the disease (e.g. susceptible, exposed, susceptible with fear, infected, removed), and identified with a different color in the picture. Individuals can move from a subpopulation to another on the network of connections among subpopulations. At each time step individuals move with a commuting rate $c_{ij}$ from subpopulation $i$ to subpopulation $j$. (B) Schematic illustration of the system's critical functionality. The system if fully functional ($CF(t) = 1$) during ordinary conditions when all the subpopulations are healthy and the number of real commuters is equal to the number of virtual commuters, i.e. $D(t) = 0$ and $C(t) = Z(t)$. After the outbreak takes place ($T_0$) the system's functionality decreases because of the disease propagation and the eventual travel reduction. Next the system starts to recover until the complete extinction of the epidemic ($T_E$) which corresponds to the time when no more infected individuals are in the system. The curves (a) and (b) represent the critical functionality of scenarios corresponding to high and low values of resilience.

**Diffusion Processes.** The edge connecting two subpopulations $i$ and $j$ indicates the presence of a flux of travelers i.e. diffusion, mobility of people. We assume that individuals in the subpopulation $i$ will visit the subpopulations $j$ with a per capita diffusion rate $d_{ij}$ on any given edge [39] (see the Methods section for further details). We define the total number of travelers $Z$ between the subpopulations $i$ and $j$ at time $t$ as $Z_{ij}(t) = d_{ij}N_i(t)$, so that when the system is fully functional, the total number of travelers at time $t$ from the node $i$ is $Z_i(t) = \sum_{j \in k_i} Z_{ij}(t)$. Under these conditions, the total number of travelers in the metapopulation system at time $t$ is simply $Z(t) = \sum_i Z_i(t)$.

**Table 1. Notation and description of the parameters used in our simulations.**

| Notation | Description |
|---|---|
| $V$ | Number of subpopulations in the metapopulation network |
| $N$ | Number of individuals in the system |
| $\langle k \rangle$ | Average degree of the metapopulation network |
| $D$ | Number of diseased populations |
| $H$ | Fraction of healthy populations |
| $A$ | Fraction of active travelers in the system |
| $p$ | The parameter that regulates the system wide travel restrictions |
| $r$ | System's resilience |
| $CF$ | System's critical functionality |
| $T_c$ | Resilience control time |
| $S$ | Susceptible individuals |
| $S^F$ | Susceptible individuals with fear |
| $E$ | Exposed individuals |
| $I$ | Infected individuals |
| $R$ | Recovered individuals |
| $R_0$ | Basic reproduction number |
| $\lambda$ | The rate at which an 'exposed' person becomes 'infected' |
| $\mu$ | The rate at which an 'infected' recovers and moves into the 'recovered' compartment |
| $\beta$ | The parameter controlling how often a 'susceptible'-'infected' contact results in a new 'exposed' |
| $\beta^F$ | The parameter controlling how often a 'susceptible'-'infected' contact results in 'susceptible individual with fear' |
| $\alpha$ | The parameter controlling how often a 'susceptible'-'susceptible individuals with fear' contact results in a new 'susceptible individual with fear' |
| $r_b$ | The parameter that modulates the level of self-induced behavioral change that leads to the reduction of the transmission rate |
| $\mu_F$ | The rate at which individual with fear moves back into the 'susceptible' compartment |

In the following we assume that infected individuals do not travel between subpopulations, thus reducing the actual number of travelers.

**Reaction Processes.** In analyzing contagion processes we extend the compartmental scheme of the basic SEIR model [40,41] (see Methods and Supplementary Information (SI) for a detailed description). Indeed an important element in the mitigation of epidemics is self-initiated behavioral changes triggered in the population by awareness/fear of the disease [42,43]. These generally reduce the transmissibility and spreading. Examples of behavioral changes include social distancing behaviors such as avoidance of public places, working from home, decrease of leisure and business travel etc. In order to include behavioral changes in our model, we consider a separate behavioral class within the population [44], defining a special compartment of susceptible individuals, $S^F$, where $F$ stands for "fearful". In particular, individuals transition to this compartment depending on the prevalence of infected and other fearful individuals according to a rate $\beta_F$. This rate mimics the likelihood that individuals will adopt a different social behavior as a result of the increased awareness of the disease as perceived from the number of infected and fearful individuals present in the system. Clearly, spontaneous or more complex types of transitions (for example indirectly linked to the disease transmission due to mass media effects[44]) could be considered. However, they would require more parameters and introduce other non-trivial dynamics. We leave the study of other behavioral changes models for future works. It follows that in each subpopulation the total number of individuals is partitioned into the compartments $S(t), S^F(t), E(t), I(t), R(t)$ denoting the number of susceptible, fearful, exposed, infected, and removed individuals at time $t$, respectively. The transition processes are defined by the following scheme: $S + I \rightarrow E + I, S + I \rightarrow S^F + I, S + S^F \rightarrow 2S^F, S^F + I \rightarrow E + I, E \rightarrow I$ and $I \rightarrow R$ with their respective reaction rates, $\beta, \beta_F, \alpha \beta_F, r_b \beta, \lambda$ and $\mu$. Analogously, individuals in the $S^F$ compartment may transition back in the susceptible compartment with a rate $\mu_F$, $S^F + S \rightarrow S$. The model reverts to the classic SEIR if $\beta_F = 0$ (the detailed presentation of the dynamic is reported in the SI). The basic reproductive number of an SEIR model is $R_0 = \beta/\mu$. This quantity determines the average number of infections generated by one infected individual in a fully susceptible population. In each subpopulation the disease transmission is able to generate a number of infected individuals larger than those who recover only if $R_0 > 1$, yielding the classic result for the epidemic threshold [45]; if the spreading rate is not large enough to allow a reproductive number larger than one (i.e., $\beta > \mu$), the epidemic outbreak will affect only a negligible portion of the population and will quickly die out (the model details are reported in the Methods section).

**System's resilience**. Here, we introduce a quantitative measure that captures and implements the definition of resilience in epidemic modelling, similarly to what proposed in Ganin et al.[32,34].

Among the many possible elements defining the resilience of a system, we consider the system-wide critical functionality as a function of the individual's risk of getting the disease and the disruption to the system's functionality generated by the human mobility deterioration. For the sake of simplicity, in our model we assume that infected individuals do not travel. The extension to models in which a fraction of infected individuals are traveling is straightforward[4] with the only effect of decreasing the timescale for the disease spreading, but not altering the overall dynamic of the system. Furthermore, as discussed below, the system might be subject to other travel limitations. As a result, during the epidemic we have an overall decrease in the mobility flows with respect to a disease-free scenario. It follows that the number of travelers between subpopulations $i$ and $j$ at time $t$ is $C_{ij}(t) = c_{ij} \tilde{N}_i(t)$, where $c_{ij}$ is the adjusted diffusion rate, $\tilde{N}_i(t) = S_i(t) + E_i(t) + R_i(t)$, and the total number of commuters in the metapopulation system at time $t$ is given by $C(t) = \sum C_i(t)$. Note that in general, $c_{ij} < d_{ij}$. This can be naturally related to a deterioration of the system-level critical functionality as it corresponds to economic and financial losses as well as logistic and infrastructural service disruption. In order to evaluate the system's loss of critical functionality related to the travel restrictions, we define the fraction of active travelers at time $t$ as $A(t) = C(t)/Z(t)$. Analogously, we characterize the system's risk related to the disease propagation as the fraction of healthy subpopulations $H(t) = 1 - D(t)/V$, where $D(t)$ is the number of diseased subpopulations at time $t$ and $V$ is the total number of subpopulations in the system. The number of diseased subpopulations accounts for the amount of risk posed to individuals in the system, which we assume to be proportional to the overall attack rate and expresses the vulnerability of the networked system [36,46,47]. Here, as the model assumes statistically equivalent subpopulations, the attack rate is proportional to the number of subpopulations affected by the epidemic. At time $t$, we define the critical functionality, $CF(t)$, (Figure 1B), as the product of the fraction of active travelers $A(t)$ and the fraction of healthy populations $H(t)$, i.e. $CF(t) = H(t) \cdot$

$A(t)$. Per our earlier definition of resilience [32] $r$, we evaluate it as the integral over time of the critical functionality, normalized over the control time $T_C$ so that $r \in [0,1]$:

$$r = \frac{1}{T_C} \int_0^{T_C} CF(t)dt. \qquad (1)$$

The control time $T_C$ corresponds to the maximum extinction time $T_E$ for different values of epidemic reproductive number $R_0$ (see the Supporting Information for further detail). Resilience, therefore, also includes the time dimension, in particular, the time to return to full functionality, as defined by the system's critical elements. In reference[32] we provided an operational definition of resilience starting from the concepts advanced by the National Academy of Sciences in USA. In this paper, we apply such general framework to the case of disease spreading. Furthermore, we extend it to reaction-diffusion processes on metapopulations. In the following, we will quantitatively characterize different containment/mitigation interventions via a critical functionality analysis. Desirable (optimal) strategies correspond to high (maximum) value of $r$. It is worth remarking that, for the sake of simplicity, we use here a definition of critical functionality that weights equally the two components $A(t)$ and $H(t)$. Thus, our findings are constrained by such choice. The two contributions could be weighted differently, i.e. $CF(t) = H(t)^\alpha \cdot A(t)^\beta$. However, our aim is to highlight the importance of going beyond "model-based" approach to epidemic analysis and move towards system resilience assessments. In this spirit, we opted for the simplest definition of critical functionality able to capture the two most used metrics in model-based approaches: epidemic risk and mobility. We used the multiplication of the two quantities because it makes the critical functionality more sensitive to small changes of the values. Furthermore, by multiplying two ratios we don't need to add a normalization factor (the critical functionality is defined in the interval $[0,1]$). In more realistic context, and depending on the precise cost-benefit analysis, the various terms may be weighted differently and more complex functional form for the critical functionality can be defined.

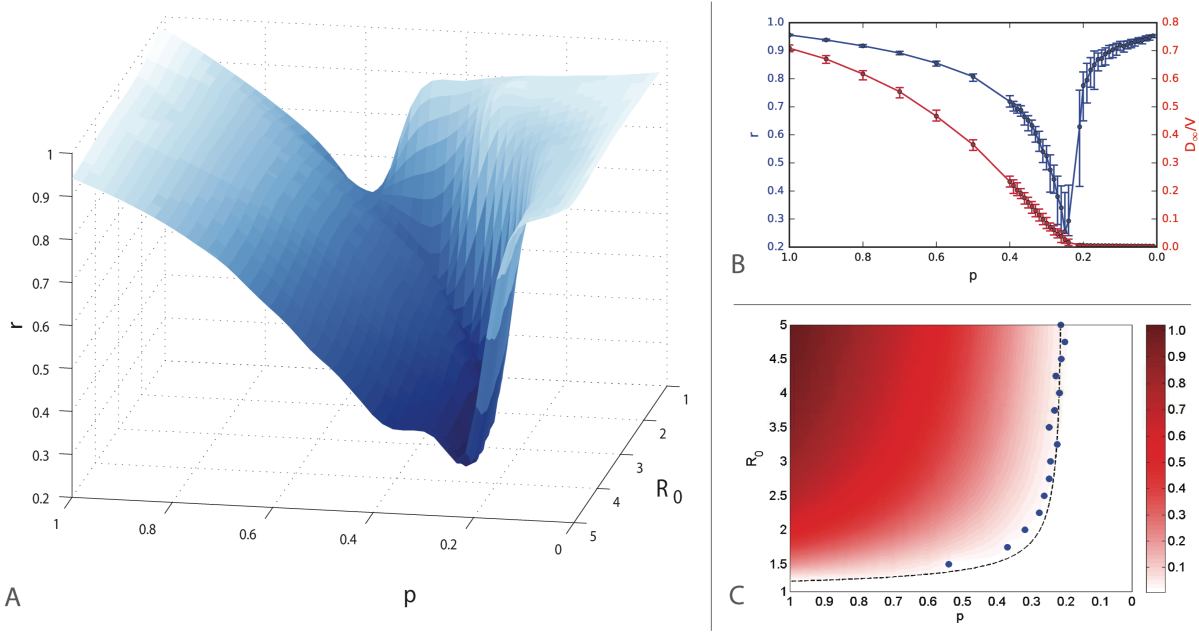

**Figure 2. Resilience and final fraction of diseased populations in the heterogeneous metapopulation system with traffic dependent diffusion rates.** (A) 3D surface representing resilience in a homogeneous metapopulation system as a function of local threshold $R_0$ and the diffusion rate $p$: the minimum value of resilience separates two regions associated to values very close to the optimal case. (B) Cross-sections (blue) of the 3D plot for $R_0 = 3.5$ and its comparison with the final fraction of diseased populations (red): while the reduction of the diffusion rate $p$ brings to a constant the fraction of diseased populations it also causes an initial decrease of resilience to a minimum value after which it starts increasing and the system returns to its optimal conditions. (C) The map of the final fraction of diseased populations $D_\infty / V$ is shown as a function of the local epidemic threshold $R_0$ and the travel diffusion $p$. We show that the minimum values of resilience (blue points) correspond to the theoretical value of the final fraction of diseased subpopulations $D_\infty / V$ at the end of the global epidemic (black line).

Among other things, these type of analysis could consider: i) the details of the disease spreading in the population such as mortality, infectiousness, recovery time, and possible residual immunity ii) the preparedness, measured in terms of availability of vaccines, anti-virals, hospital beds, or intensive care units, iii) the socio-economical costs induced by a major outbreak and by interventions such as travel bans, school closures etc. iv) politics and public perception of risk.

**Effects of system-wide travel restrictions.** Epidemic containment measures, based on limiting or constraining human mobility, are often considered in the contingency planning and always re-emerge when there are new infectious disease threats [1]. The target of these control measures are travels to/from the areas affected by an epidemic outbreak and the corresponding decrease of infected individuals reaching areas not yet affected by the epidemic. At the same time, travel restrictions have a large impact on the economy and affect the delivery of services, including medical supplies and the deployment of specialized personnel to manage the epidemic. For this reason, travel restrictions must be carefully scrutinized to trade off the costs and benefits. We introduce the parameter $p \in [10^{-5}, 1]$ that allows us to simulate policy-induced system-wide travel restrictions. In our settings, such measures are active until the disease is circulating in the system, i.e. there is at least one infected individual across all subpopulations. In the case of no travel restrictions and/or after the disease dies out, we have $p = 1$. In the case of travel restrictions ($p < 1$), we rescale travel flow so that mobility is a fraction of that in the unaffected system; i.e. $c_{ij} = p \cdot d_{ij}$. To better understand the effect of such mitigation strategy, let us consider the classic SEIR model by setting $\beta_F = 0$. In the presence of travel restrictions and depending on the level of mixing, each subpopulation may or may not transmit the infection or contagion process to another subpopulation it is in contact with. In other words, the mobility parameter $p$ influences the probability that exposed individuals will export the contagion process to other regions of the metapopulation network. Further, it introduces a transition between a regime in which the contagion process may invade a macroscopic fraction of the network and a regime in which it is limited to a few subpopulations. The transition is mathematically characterized by the global invasion threshold $R_*$ [45]. This is the analogue of the basic reproductive number at the subpopulations level and defines the average number of infected subpopulations generated by one infected subpopulation in a fully susceptible metapopulation system. In general, $R_*$ is a function of the basic epidemic parameters, including $R_0$, and the mobility parameter $p$. The invasion threshold occurs at the critical value $p_c$ for which $R_* = 1$. In some cases, $p_c$ can be evaluated analytically (see the *Methods* section). In general, it can be estimated numerically by measuring the number of infected subpopulations as a function of the parameter $p$. Risk, as measured in terms of attack rate, is, therefore, monotonically decreasing due to increasingly restricted travel, and falls to virtually zero for values of $p$ below the invasion threshold. Thus, from a risk perspective, the best strategy during a disease outbreak is to reduce the mobility. However, an inspection of the profile of resilience provides a different picture. In Figure 2 we report the value of $r$ obtained by sampling the phase space of the model $p - R_0$ for different values of the travel diffusion parameter and the epidemic reproductive number in heterogeneous metapopulation systems (a comparison between homogenous and heterogeneous networks is reported in SI). Each point of the phase space is studied by performing 100 stochastic realizations. The 3D dimensional plot in the $p, R_0, r$ space reported in Figure 2A indicates that the overall resilience profile is characterized by a sharp drop as we approach the invasion threshold, i.e. $p \to p_c$. Figure 2B shows that, while the risk decreases, the reduction of the diffusion rate $p$ causes a reduction of $r$ until the global invasion threshold, after which the resilience value rapidly increases. This effect is mainly due to the critical slowing of the epidemic spreading near the invasion threshold. Indeed, close to the threshold, the epidemic is still in a supercritical state, but it takes increasingly longer time to invade the system as the threshold is approached. This can be simply related to the divergence of the invasion doubling time $T_d$, which is defined as the time until the number of infected subpopulations doubles, relative to that at some other time. The doubling time is related to the subpopulation reproductive number as $T_d \sim (R_* - 1)^{-1}$, leading to a divergence of the doubling time as the invasion threshold is approached for $R_* \to 1$. Although the absolute risk is very low, the system remains in a state of deteriorated functionality (restrictions in travels) for longer and longer times [48]. The decrease of functionality is not offset by a corresponding decrease of risk, and the minimum in resilience is attained exactly at the global invasion threshold. The comparison between the theoretical values of the invasion threshold and the computed minimum values of resilience is reported in Figure 2C.

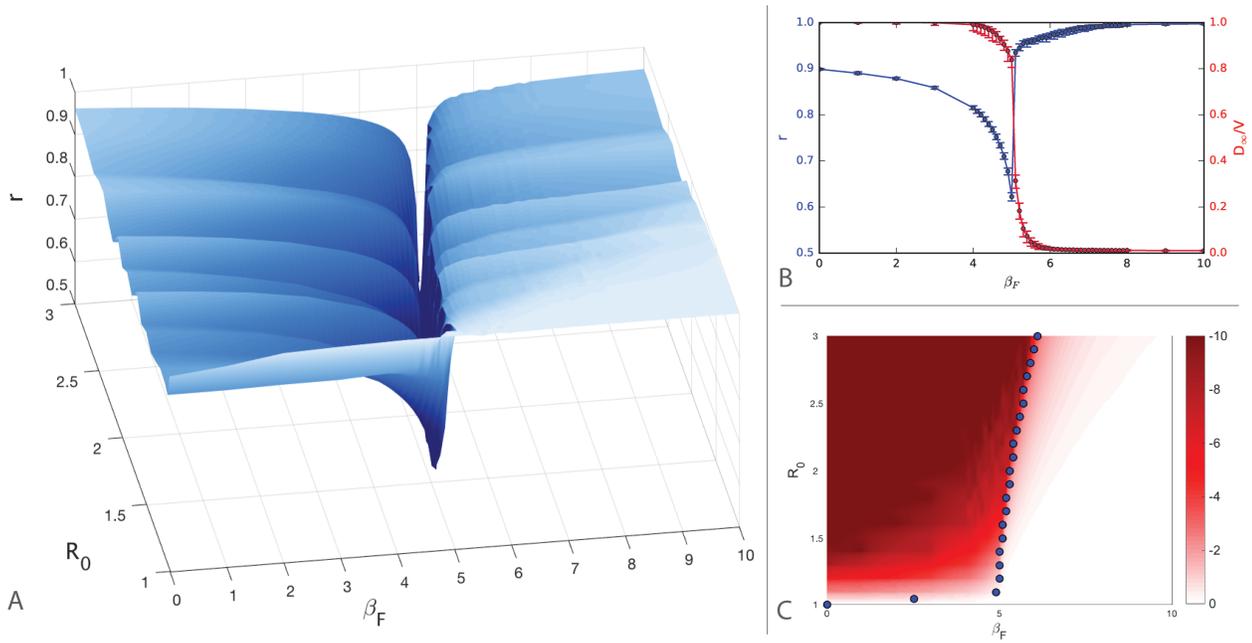

**Figure 3. Resilience and diseased populations in a heterogeneous metapopulation system with individual self-dependent travel reduction.** (A) 3D surface representing resilience in a heterogeneous metapopulation system as a function of local threshold $R_0$ and the fear parameter $\beta_F$: two areas of high values of resilience are separated with a narrow region of very low ones. (B) Comparison between resilience (blue) reported as cross-sections of the 3D plot for $R_0 = 1.3$ and the final fraction of diseased populations $D_\infty/V$ (red): while the increase of the fear transmissibility parameter $\beta_F$ brings to a constant the fraction of the diseased populations it also causes an initial decrease of resilience to a minimum value after which the system bounces back to optimal conditions. (C) Even in this case the minimum values of resilience (blue points) correspond to the transition region from high to low final diseased populations. The colormap of the logarithmic of the healthy populations ($log(1 - D_\infty/V)$) is shown as a function of the local epidemic threshold $R_0$ and the fear parameter $\beta_F$.

**Effects of self-initiated behavioural changes.** In order to isolate the effects of behavioural changes, in this section the travel parameter is kept constant with $p = 1$. Individuals in the $S^F$ compartment adopt travel avoidance so that $\beta_F$ plays a similar role to the travel restriction as reported in Figure 3. Furthermore, inside each subpopulation, individuals in the $S^F$ compartment reduce their contacts, thus decreasing the likelihood to become infected. Overall, the presence of self-initiated behavioral changes in a population results in a reduction of the final epidemic size. In this setting, we have explored a phase space of parameters constituted of $R_0 \in [1.01, 3]$ and $\beta_F \in$

[0,20] (see the *Methods* section for the other model parameters). In Figure 3A we quantify resilience for different values of the fear parameter $\beta_F$ in heterogeneous metapopulation systems. The 3D dimensional plot in the $\beta_F, R_0, r$ space shows a clear similarity with the travel restrictions scenario. Figure 3B shows that, while increasing $\beta_F$ leads to a decrease in risk, it also induces a reduction of resilience. It is possible to observe that, even in this case, the minimum values of $r$ are related to the invasion threshold. In Figure 3C the phase diagram of the fraction of diseased populations at the end of the simulations $D_\infty/V$ is reported in the $\beta_F, R_0$ space. This picture shows that there is a critical value of the fear transmissibility parameter $\beta_F$, after which the fraction of diseased populations $D_\infty/V$ starts to decrease (i.e. $D_\infty/V < 1$). The minimum value of resilience, in this case, corresponds to the value of the fear transmissibility, after which a reduction of the fraction of diseased populations is observed. Although the approach to this critical boundary corresponds to a reduction of the infection risk, similarly to the case of travel restrictions, the measured resilience of the system decreases and attains its minimum value right at the transition point.

**Effects of system-wide travel restrictions in data-driven simulations.** As a further confirmation of the validity of the theoretical construct above described, we tested our results in a data-driven modelling setting. We considered the Global Epidemic and Mobility model (GLEAM) [3,49] which integrates high resolution demographic and mobility data by using a high-definition, geographically structured metapopulation approach. The model's technical details and the algorithms underpinning the computational implementation have been extensively reported in the literature. GLEAM is a spatial, stochastic and individual-based epidemic model that divides the world population into geographic census areas, defined around transportation hubs and connected by mobility fluxes. The population of each census area is obtained by integrating data from the high-resolution population database of the 'Gridded Population of the World' project of the Socioeconomic Data and Application Center at Columbia University (SEDAC).

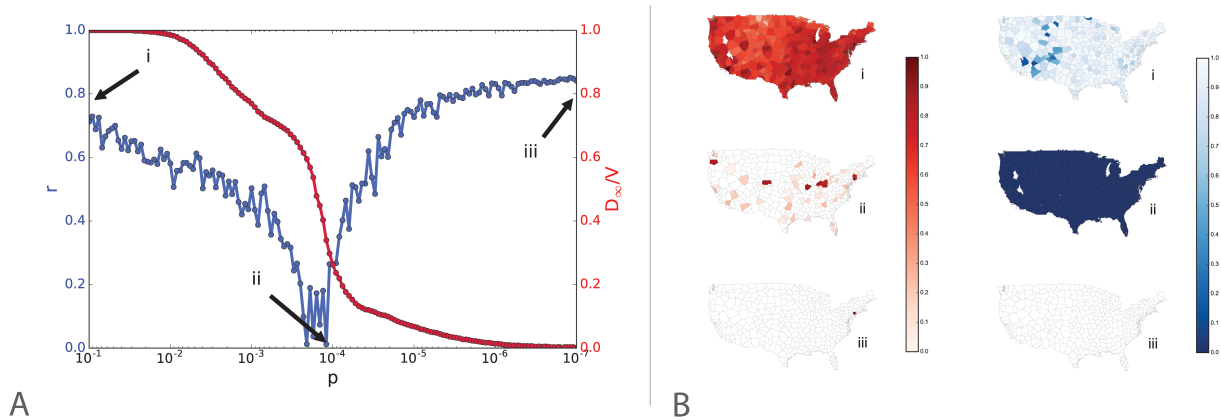

**Figure 4. Resilience and epidemic size in the data-driven scenario.** (A) The plot shows the difference between resilience (blue) and the final fraction of diseased populations (red) for different values of the diffusion rate $p$. Here, we can identify three critical regions of the system. i) diffusion rate $p = 0.1$ above the critical invasion threshold. Even if the system is characterized by sub-optimal resilience, the disease spreads all over the system. ii) the reduction of the diffusion parameter $p$ results in a significant decrease of the number of diseased populations but also in a dramatic decrease of resilience; iii) below the critical invasion threshold resilience goes back to high values as fraction of diseased populations approaches zero. (B) Epidemic size (red) and resilience (blue) for the different values of the diffusion parameter $p$ corresponding to the three aforementioned regions. Python 2.7 (https://www.python.org/) and the Basemap library (https://pypi.python.org/pypi/basemap/1.0.7) were used to create these maps.

The mobility among subpopulations is comprised of global air travel and the small-scale movement between adjacent subpopulations; i.e., the daily commuting patterns of individuals. Commuting and short-range mobility considers data from 80,000 administrative regions in 5 different continents. Here, we considered the Continental United States and simulated an SEIR contagion process, in which infected individuals do not travel. The number of infected subpopulations at the end of an outbreak and resilience as a function of the global mobility restrictions that result are shown in Figure 4. The initial conditions of the epidemic were set with 5 infected individuals in the city of New York, assuming $\beta = 0.48$, $\lambda = 0.66$ and $\mu = 0.45$. Mobility restrictions are implemented by reducing all the mobility flows connecting diseased subpopulations by a factor $p$, thus considering the heterogeneities of the subpopulations due to their different local mobility patterns (see SI). The control time $T_C$ used in the calculation of $r$ corresponds to the epidemic extinction time for the different values of the diffusion rate.

As with the theory-driven model here we observe that a reduction of the travel diffusion $p$ brings a constant reduction of diseased populations, but also reduces resilience until a critical value $p_c = 1.2 \cdot 10^{-4}$. In Figure 4B we illustrate the geographical spreading of the contagion process and the reduction of traveling of each subpopulation tracked by the model in the Continental USA for values of $p$ corresponding to three different regions of the diagram of Figure 4A. The figure clearly illustrates three regimes: i) for low travel reduction, a very severe epidemic hits all the subpopulations, but the short duration allows the system to go back to normal in a short time (high values of resilience); ii) for travel reduction close to the global invasion threshold, a small number of subpopulations are hit but the system critical functionality is compromised for a very long time, thus, resulting in a low values of resilience; iii) travel reduction above the critical threshold allows the system to contain the epidemic at the origin with low risk and high values of resilience. It is worth remarking that in the data-driven model, the minimum value of resilience is reached for travel restrictions that correspond to a reduction of mobility of three to four orders of magnitude. This is because in modern transportation networks the global invasion threshold, as already pointed out in other studies [10,14,20,39], is reached only for very severe travel restrictions that are virtually impossible to achieve. In other words, in realistic settings the feasible increase of travel restrictions appears always to decrease resilience. This calls for a careful scrutiny of the trade-off between individual's risk and resilience, as the region where both are achieved is virtually not accessible.

**Discussion**

The realistic threat quantification is difficult and evaluation of vulnerabilities and consequences of new disease epidemics is certainly a challenge. We analyzed the impact of an infectious disease epidemic in structured populations by considering a definition of system resilience that takes into consideration not only the number of infected individuals but also society's need for maintaining certain critical functions in space and time[37]. In particular, we assume that the limitations and disruptions of human mobility deteriorate the system's functionality. We observe that containment measures, that limit individuals' mobility, are advantageous in reducing risk but may deteriorate the system's functionality for a very long time and thus correspond to low resilience. Although we have considered only two of the many dimensions encompassing the functionality of socio-technical systems [28,30], we show that study of resilience allows stakeholders to measure the impact of epidemic threats and differentiate between different management alternatives. It is straightforward to envision more realistic definition of the critical functionality. The components of critical functionality could be weighted according to objective/subjective evaluation of their relevance to stakeholders. Finally, cost-benefit analyses and ethical considerations should be included in the analysis of the societal impacts of disease that could lead to long lasting effects or even death of the affected individuals. This study highlights the importance of resilience-focused analysis for selecting intervention strategies. The natural tendency to be conservative in managing potentially inflated risks associated with new and emerging epidemics can result in unnecessary burdensome and possibly ineffective actions like quarantines as well as travel bans[50]. The emerging field of resilience assessment and management[29] and its implementation[32,34,35] could thus evaluate cross-domain alternatives to identify a policy design that enhances the system's ability to (i) plan for such adverse events, (ii) absorb stress, (iii) recover, and (iv) predict and prepare for future stressors through necessary adaptation. To this end, the framework we presented can be of potential use for optimizing the policy response to a disease outbreak by balancing risk reduction with the disruption to critical functions that is associated with public health interventions.

**Methods**

**Disease propagation and self-initiated behavioral changes**. The metapopulation system is described by a scale-free network (SF) with a power-law degree distribution $P(k) \sim k^{(-\gamma)}$, which is generated by the configuration model[51] with the minimum degree $m = 2$, $\gamma = 2.1$. (For the travel restriction scenario, in the SI, we report a comparison of the results between the heterogeneous networked system described above and a metapopulation system formed by a random network with Poisson degree distribution, which is generated by the Erdos–Rényi (ER) model[52]). The networks have $V = 5000$ nodes and average degree $\langle k \rangle \sim 6$, while the total number of individuals is $N = V^2 = 25 \cdot 10^6$ which are distributed among the subpopulations nodes proportional to their degree distribution. At the beginning, 10 populations are selected at random and 50 individuals are set as exposed. All other individuals across the system are initially susceptible. We study a compartmental scheme that extends the basic SEIR[40] model by considering separate behavioral classes within populations (see SI for the detailed description of the model). For this reason, we assume that individuals can spontaneously change their behavior because of the fear of the disease entering in a specific compartment $S^F$ of susceptible individuals. In the case of travel restrictions, we set the transition rate from exposed to infected $\lambda = 0.67 \ days^{-1}$ and recovery rate $\mu = 0.33 \ days^{-1}$. In the case of the behavioral model, we set the disease parameters $\lambda = 0.3 \ days^{-1}$ and $\mu = 0.1 \ days^{-1}$ while we consider an infection probability reduction $r_b = 0.15$, the self-reinforcement parameter $\alpha = 0.1$ and the relaxation parameter $\mu_F = 0.5$. All the presented results are averaged over 100 simulations.

**Mobility process.** We adopt a simplified mechanistic approach that uses a Markovian assumption for modeling migration among subpopulations; at each time step, the movement of individuals is given according to a matrix $d_{ij}$ that expresses the probability that an individual in the subpopulation $i$ is traveling to the subpopulation $j$. We assume that the diffusion rate on any given edge from a subpopulation node of degree $k_i$ to a subpopulation node of degree $k_j$ is proportional to $k_j$ [39] and it is given by $d_{ij} = \omega_0 \left(k_i k_j\right)^\theta / T_i$, where $T_i = \sum_j w_{ij} = \sum_j \omega_0 \left(k_i k_j\right)^\theta$ represents the total flow in $i$, while $\theta$ and the exponent $\omega_0$ are system specific (e.g., and $\theta = 0.5$ and in the world-wide air transportation network [53]). In this scenario, we consider $\theta = 0.5$ and $\omega_0 = 10^{-3}$.

**Global invasion threshold.** For the SEIR model it is possible to explicitly calculate the average number of infected subpopulations for each infected subpopulation in a fully susceptible metapopulation system as $R_* = \widehat{N} p \frac{2(R_0-1)^2}{(\mu+\lambda)R_0^2} \frac{\langle k^2 \rangle - \langle k \rangle}{\langle k \rangle^2}$ [3] where $\widehat{N}$ represents the average number of individuals in a subpopulation. The condition $R_* = 1$ defines the invasion threshold for the system. Only for $R_* > 1$ can the epidemic spread to a large number of subpopulations. The invasion threshold readily provides an explicit condition for the critical mobility $p_c$, below which an epidemic cannot invade the metapopulation system, yielding the equation $p_c = \frac{1}{\widehat{N}} \frac{\langle k \rangle^2}{\langle k^2 \rangle - \langle k \rangle} \frac{(\mu+\lambda)R_0^2}{2(R_0-1)^2}$ [39].

## References and Notes

1. Riley, S. Large-scale spatial-transmission models of infectious disease. *Science* **316,** 1298–1301 (2007).
2. Marathe, M. & Vullikanti, A. K. S. Computational epidemiology. *Commun. ACM* **56,** 88–96 (2013).
3. Balcan, D. *et al.* Seasonal transmission potential and activity peaks of the new influenza A (H1N1): a Monte Carlo likelihood analysis based on human mobility. *BMC Med.* **7,** 1 (2009).
4. Balcan, D. *et al.* Multiscale mobility networks and the spatial spreading of infectious diseases. *Proc. Natl. Acad. Sci.* **106,** 21484–21489 (2009).
5. Grais, R. F., Ellis, J. H. & Glass, G. E. Assessing the impact of airline travel on the geographic spread of pandemic influenza. *Eur. J. Epidemiol.* **18,** 1065–1072 (2003).
6. Colizza, V., Barrat, A., Barthelemy, M., Valleron, A.-J. & Vespignani, A. Modeling the worldwide spread of pandemic influenza: baseline case and containment interventions. *PLoS Med* **4,** e13 (2007).
7. Hufnagel, L., Brockmann, D. & Geisel, T. Forecast and control of epidemics in a globalized world. *Proc. Natl. Acad. Sci. U. S. A.* **101,** 15124–15129 (2004).
8. Eubank, S. *et al.* Modelling disease outbreaks in realistic urban social networks. *Nature* **429,** 180–184 (2004).
9. Merler, S. & Ajelli, M. The role of population heterogeneity and human mobility in the spread of pandemic influenza. *Proc. R. Soc. Lond. B Biol. Sci.* **277,** 557–565 (2010).
10. Cooper, B. S., Pitman, R. J., Edmunds, W. J. & Gay, N. J. Delaying the international spread of pandemic influenza. *PLoS Med* **3,** e212 (2006).

11. Ferguson, N. M. *et al.* Planning for smallpox outbreaks. *Nature* **425,** 681–685 (2003).

12. Ferguson, N. M. *et al.* Strategies for mitigating an influenza pandemic. *Nature* **442,** 448–452 (2006).

13. Germann, T. C., Kadau, K., Longini, I. M. & Macken, C. A. Mitigation strategies for pandemic influenza in the United States. *Proc. Natl. Acad. Sci.* **103,** 5935–5940 (2006).

14. Epstein, J. M. *et al.* Controlling pandemic flu: the value of international air travel restrictions. *PloS One* **2,** e401 (2007).

15. Degli Atti, M. L. C. *et al.* Mitigation measures for pandemic influenza in Italy: an individual based model considering different scenarios. *PloS One* **3,** e1790 (2008).

16. Team, W. E. R. Ebola virus disease in West Africa—the first 9 months of the epidemic and forward projections. *N Engl J Med* **2014,** 1481–1495 (2014).

17. Meltzer, M. I. *et al.* Estimating the future number of cases in the Ebola epidemic—Liberia and Sierra Leone, 2014–2015. *MMWR Surveill Summ* **63,** 1–14 (2014).

18. Rivers, C. M., Lofgren, E. T., Marathe, M., Eubank, S. & Lewis, B. L. Modeling the Impact of Interventions on an Epidemic of Ebola in Sierra Leone and Liberia. *PLOS Curr. Outbreaks* (2014).

19. Shaman, J., Yang, W. & Kandula, S. Inference and forecast of the current West African Ebola outbreak in Guinea, Sierra Leone and Liberia. *PLOS Curr. Outbreaks* (2014).

20. Gomes, M. F. *et al.* Assessing the international spreading risk associated with the 2014 West African Ebola outbreak. *PLOS Curr. Outbreaks* (2014).

21. Van Kerkhove, M. D. & Ferguson, N. M. Epidemic and intervention modelling: a scientific rationale for policy decisions? Lessons from the 2009 influenza pandemic. *Bull. World Health Organ.* **90,** 306–310 (2012).

22. Lipsitch, M., Finelli, L., Heffernan, R. T., Leung, G. M. & Redd; for the 2009 H1N1 Surveillance Group, S. C. Improving the evidence base for decision making during a pandemic: the example of 2009 influenza A/H1N1. *Biosecurity Bioterrorism Biodefense Strategy Pract. Sci.* **9,** 89–115 (2011).

23. Van Kerkhove, M. D. *et al.* Studies needed to address public health challenges of the 2009 H1N1 influenza pandemic: insights from modeling. *PLoS Med* **7,** e1000275 (2010).

24. Merler, S. *et al.* Spatiotemporal spread of the 2014 outbreak of Ebola virus disease in Liberia and the effectiveness of non-pharmaceutical interventions: a computational modelling analysis. *Lancet Infect. Dis.* **15,** 204–211 (2015).

25. Wu, J. T., Riley, S., Fraser, C. & Leung, G. M. Reducing the impact of the next influenza pandemic using household-based public health interventions. *PLoS Med* **3,** e361 (2006).

26. Organization, W. H. & others. WHO guide to identifying the economic consequences of disease and injury. (2009).

27. DeWitte SN, A. C., Kurth MH & I, L. Disease epidemics: lessons for resilience in an increasingly connected world. *J. Public Health* (2016).

28. Holling, C. S. Resilience and stability of ecological systems. *Annu. Rev. Ecol. Syst.* 1–23 (1973).

29. Linkov, I. *et al.* Measurable resilience for actionable policy. *Environ. Sci. Technol.* **47,** 10108–10110 (2013).

30. Linkov, I. *et al.* Changing the resilience paradigm. *Nat. Clim. Change* **4,** 407–409 (2014).

31. NATO Advanced Research Workshop on Resilience-Based Approachers to Critical Infrastructure Safegaurding, Linkov, I. & Palma-Oliveira, J. M. *Resilience and risk: methods and application in environment, cyber and social domains*. (2017).

32. Ganin, A. A. *et al.* Operational resilience: concepts, design and analysis. *Sci. Rep.* **6,** (2016).

33. Barrett, C. B. & Constas, M. A. Toward a theory of resilience for international development applications. *Proc. Natl. Acad. Sci.* **111,** 14625–14630 (2014).

34. Ganin, A. A. *et al.* Resilience and efficiency in transportation networks. *Sci. Adv.* **3,** e1701079 (2017).

35. Gao, J., Barzel, B. & Barabási, A.-L. Universal resilience patterns in complex networks. *Nature* **530,** 307–312 (2016).

36. Majdandzic, A. *et al.* Spontaneous recovery in dynamical networks. *Nat. Phys.* **10,** 34–38 (2014).

37. Linkov, I., Fox-Lent, C., Keisler, J., Della Sala, S. & Sieweke, J. Risk and resilience lessons from Venice. *Environ. Syst. Decis.* **34,** 378–382 (2014).

38. Lu, D., Yang, S., Zhang, J., Wang, H. & Li, D. Resilience of epidemics for SIS model on networks. *Chaos Interdiscip. J. Nonlinear Sci.* **27,** 083105 (2017).

39. Colizza, V. & Vespignani, A. Epidemic modeling in metapopulation systems with heterogeneous coupling pattern: Theory and simulations. *J. Theor. Biol.* **251,** 450–467 (2008).

40. Anderson, R. M., May, R. M. & Anderson, B. *Infectious diseases of humans: dynamics and control*. **28,** (Wiley Online Library, 1992).

41. Keeling, M. J. & Rohani, P. *Modeling infectious diseases in humans and animals*. (Princeton University Press, 2008).

42. Funk, S., Salathé, M. & Jansen, V. A. Modelling the influence of human behaviour on the spread of infectious diseases: a review. *J. R. Soc. Interface* rsif20100142 (2010).

43. Massaro, E. & Bagnoli, F. Epidemic spreading and risk perception in multiplex networks: a self-organized percolation method. *Phys. Rev. E* **90,** 052817 (2014).

44. Perra, N., Balcan, D., Gonçalves, B. & Vespignani, A. Towards a characterization of behavior-disease models. *PloS One* **6,** e23084 (2011).

45. Colizza, V. & Vespignani, A. Invasion threshold in heterogeneous metapopulation networks. *Phys. Rev. Lett.* **99,** 148701 (2007).

46. Buldyrev, S. V., Parshani, R., Paul, G., Stanley, H. E. & Havlin, S. Catastrophic cascade of failures in interdependent networks. *Nature* **464,** 1025–1028 (2010).

47. Gao, J., Buldyrev, S. V., Havlin, S. & Stanley, H. E. Robustness of a network of networks. *Phys. Rev. Lett.* **107,** 195701 (2011).

48. Fraser, C., Riley, S., Anderson, R. M. & Ferguson, N. M. Factors that make an infectious disease outbreak controllable. *Proc. Natl. Acad. Sci. U. S. A.* **101,** 6146–6151 (2004).

49. Tizzoni, M. *et al.* Real-time numerical forecast of global epidemic spreading: case study of 2009 A/H1N1pdm. *BMC Med.* **10,** 1 (2012).

50. Hayden, C. & others. The ebola questions. *Nature* **514,** 554–557 (2014).

51. Catanzaro, M., Boguná, M. & Pastor-Satorras, R. Generation of uncorrelated random scale-free networks. *Phys. Rev. E* **71,** 027103 (2005).

52. Erdős, P. & Rényi, A. On the evolution of random graphs.

53. Barrat, A., Barthelemy, M., Pastor-Satorras, R. & Vespignani, A. The architecture of complex weighted networks. *Proc. Natl. Acad. Sci.* **101,** 3747–3752 (2004).

## Acknowledgments

The views and opinions expressed in this article are those of the individual authors and not those of the US Army, and other sponsor organizations. This study was supported by the U.S. Army Engineer Research and Development Center (Dr. E. Ferguson, Technical Director) and by the Defense Threat Reduction Agency, Basic Research Program (Dr. P. Tandy, program manager). We are grateful to Margaret Kurth for her helpful comments and assistance. AV acknowledges the support of NSF CMMI-1125095, MIDAS-National Institute of General Medical Sciences U54GM111274 awards. The authors declare no competing financial interests.

# Authors Contributions

The authors contributed to (A) conceive and design the experiments, (B) perform the experiments, (C) write the paper, (D) develop the model, (E) perform the data driven simulations and (F) analyse the data.

Emanuele Massaro A, B, C, D, E, F; Alexander Ganin A, B, C, D; Nicola Perra A, B, C, D, E, F; Igor Linkov C, D; Alessandro Vespignani A, C, D.

# Resilience management during large-scale epidemic outbreaks
Supporting Information

**Emanuele Massaro**[a,b,c], **Alexander Ganin**[a,d], **Nicolar Perra**[e,f,g], **Igor Linkov**[a], **and Alessandro Vespignani**[f,g,h]

[a]U.S. Army Corps of Engineers – Engineer Research and Development Center, Environmental Laboratory, Concord, MA, 01742, USA; [b]Senseable City Laboratory, Massachusetts Institute of Technology, 77 Massachusetts Avenue, Cambridge, MA 02139, USA; [c]HERUS Lab, École Polytechinque Fédérale de Lausanne (EPFL), CH-1015 Lausanne, Switzerland; [d]University of Virginia, Department of Systems and Information Engineering, Charlottesville, VA, 22904, USA; [e]Business School of Greenwich University, London, UK; [f]Laboratory for the Modeling of Biological and Socio-Technical Systems, Northeastern University, Boston, MA 02115, USA; [g]Institute for Scientific Interchange, 10126 Torino, Italy; [h]Institute for Quantitative Social Sciences at Harvard University, Cambridge, MA 02138, USA

**Reaction process**

The SEIR model [1] is customarily used to describe the progression of acute infectious diseases, such as influenza in closed populations, where the total number of individuals in the population is partitioned into the compartments $S(t)$, $E(t)$, $I(t)$ and $R(t)$, denoting the number of susceptible, exposed, infected and recovered individuals at time t, respectively. By definition it follows that $N(t) = S(t) + E(t) + I(t) + R(t)$. In the SEIR model we have three transitions:

$$S + I \xrightarrow{\beta} E + I$$
$$E \xrightarrow{\lambda} I \qquad [1]$$
$$I \xrightarrow{\mu} R$$

The first one, denoted by $S \to E$, is when a susceptible individual interacts with an infectious individual and enters in the exposed state with probability $\beta$. After a time period (the so-called intrinsic incubation time) $t_i = 1/\lambda$ the exposed individual becomes infected. An infected individual recovers from the disease in the viremic time $t_e = 1/\mu$. The crucial parameter in the analysis of single population epidemic outbreaks is the basic reproductive number $R_0$, which counts the expected number of secondary infected cases generated by a primary infected individual, given by $R_0 = \beta/\mu$. Here we propose a characterization of a set of prototypical mechanisms for self-initiated social distancing induced by local prevalence-based information available to individuals in the population. We characterize the effects of these mechanisms in the framework of a compartmental scheme that enlarges the basic SEIR model by considering separate behavioral classes within the population (2). In particular the fear of the disease is what induces behavioral changes in the population (3). For this reason we will assume that individuals affected by the fear of the disease will be grouped in a specific compartment SF of susceptible individuals. We consider a mechanism for which people can acquire fear assuming that susceptible individuals will adopt behavioral changes only if they interact with infectious individuals in the same subpopulations. This implies that the larger the number of sick and infectious individuals among one populations, the higher the probability for the individuals that resides in that nodes to adopt behavioral changes induced by awareness/fear of the disease. Moreover we consider the scenario in which we also consider self-reinforcing fear spread which accounts for the possibility that individuals might enter the compartment simply by interacting with people in this compartment: fear generating fear. In this model people could develop fear of the infection both by interacting with infected persons and with people already concerned about the disease. A new parameter $\alpha \geq 0$, is necessary to distinguish between these two interactions. We assume that these processes, different in their nature, have different rates. To differentiate them we consider that people who contact infected people are more likely to be scared of the disease than those who interact with fearful individuals. For this reason we set $0 \leq \alpha \leq 1$. The fear contagion process therefore can be modeled as:

$$S + I \xrightarrow{\beta_F} S^F + I \qquad [2]$$

where in analogy with the disease spread, $\beta_F$ is the transmission rate of the awareness/fear of the disease. In addition to the local prevalence-based spread of the fear of the disease, in this case we assume that the fear contagion may also occur by contacting individuals who have already acquired fear/awareness of the disease. In other words, the larger the number of individuals who have fear/awareness of the disease among one individual's contacts, the higher the probability of that individual adopting behavioral changes and moving into the class $S^F$. The fear contagion therefore can also progress according to the following process:

$$S + S_F \xrightarrow{\alpha \beta_F} 2 S^F \qquad [3]$$

Then we consider the fact that people with fear have less probability to become infected:

$$S^F + I \xrightarrow{r_b \beta} E + I \qquad [4]$$

with $0 \leq r_b < 1$ (i.e. $r_b\beta < \beta$). Moreover we consider the fact that our social behavior is modified by our local interactions with other individuals on a much more rapidly acting time-scale. The fear/awareness contagion process is not only defined by the spreading of fear from individual to individual, but also by the process defining the transition from the state of fear of the disease back to the regular susceptible state in which the individual relaxes the adopted behavioral changes and returns to regular social behavior. We can therefore consider the following processes:

$$S_F + S \xrightarrow{\mu_F} 2S \quad [5]$$

and

$$S_F + R \xrightarrow{\mu_F} S + R \quad [6]$$

Finally the system can be described by the following set of equations:

$$\begin{aligned}
d_t S(t) &= -\beta S(t)\frac{I(t)}{N} - \beta_F S(t)\left[\frac{I(t) + \alpha S^F(t)}{N}\right] + \mu_F S(t)\left[\frac{S(t) + R(t)}{N}\right] \\
d_t S^F(t) &= -r_b\beta S^F(t)\frac{I(t)}{N} + \beta_F S(t)\left[\frac{I(t) + \alpha S^F(t)}{N}\right] - \mu_F S(t)\left[\frac{S(t) + R(t)}{N}\right] \\
d_t E(t) &= -\lambda E(t) + \beta S(t)\frac{I(t)}{N} + r_b\beta S^F\frac{I(t)}{N} \\
d_t I(t) &= -\mu I(t) + \lambda E(t) \\
d_t R(t) &= \mu I(t)
\end{aligned} \quad [7]$$

The system described by the Equation 7 is reduced to classic SEIR for $\beta_F = 0$.

## Definition of the control time $T_C$

We set the control time $T_C$ as function of the epidemic extinction time $T_E$ for the different model parameters we considered. The control time $T_C$ corresponds to the maximum extinction time $T_E$ for different values of epidemic reproductive number $R_0$ an be defined.

Fig. S1 shows the epidemic extinction time decreasing the diffusion parameter p for three different values of the epidemic reproduction number $R_0$. Fig. S1 shows the value of the control time $T_C(R_0)$ used in our experiments in both homogeneous and heterogeneous networks in the diffusion case.

## Critical Thresholds

For the SEIR model identify model a critical mobility value $p_c$, below which the epidemics cannot invade the metapopulation system given by the equation [2]:

$$p_c = \frac{1}{\hat{N}}\frac{\langle k\rangle^2}{\langle k^2\rangle - \langle k\rangle}\frac{(\mu + \lambda)R_0^2}{2(R_0 - 1)^2} \quad [8]$$

where $\hat{N}$ represents the average number of individuals in a population. In Fig. S3 we report the minimum valued of the resilience (points) and the theoretical values of the invasion threshold (dotted lines) in both homogenous and heterogeneous networks. The effect of the heterogeneity on the invasion threshold in metapopulation has been previously extensively analysed [3]. In Fig. S3, Fig. S4, Fig. S5 we report the comparison between homogeneous and heterogeneous cases. Here we consider $\lambda = 0.3$ and $\mu = 0.1$.

## Self-initiated behavioral changes

Even in this scenario we observed the presence of a critical value of the precaution level $\beta_F$ after which there is the reduction of the risk in the system (see Figure 3 in the main text). In correspondence of this critical point it is possible to observe non trivial patterns of the system's functionality [4–14]. Indeed the behavioral changes though complicates the dynamics of the model [15]: in particular, within several regions of the parameter space we observe two or more epidemic peaks that produce non-trivial patterns of the system's critical functionality as shown in Fig. S6. This non-trivial behavior can be easily understood. Behavioral change is a self-reinforcing mechanism until it causes a decline in new cases. At this point individuals are lured into a false sense of security and return back to their normal behavior often causing a multiple epidemic peaks as reported in Fig. S7. Some authors believe that a similar process occurred during the 1918 pandemic, resulting in multiple epidemic peaks [16, 17]. In this following example it is possible to observe that before the critical ($\beta_F = 4.3$) point even if all the populations are interested by the disease the extinction time of the disease itself it is lower if compared with the extinction time caused by the multiple peaks caused by the increasing of the precaution level ($\beta_F = 4.3$). However after the transition point the system starts to recover fast also reducing the risk.

**Data-driven simulations: GLEAM**

In order to validate the theoretical framework developed, we considered data-driven simulations using the Global Epidemic And Mobility Model (GLEAM) [18]. GLEAM is based on three different data layers (see Ref. [18] for details). In particular,

- The population layer is based on the high-resolution population database of the Gridded Population of the World project by the Socio-Economic Data and Applications Center (SEDAC) that estimates population with a granularity given by a lattice of cells covering the whole planet at a resolution of $15x15$ minutes of arc.

- Mobility Layer integrates short-range and long-range transportation data. Long-range air travel mobility is based on travel flow data obtained from the International Air Transport Association (IATA) and the Official Airline Guide (OAG) databases, which contain the list of worldwide airport pairs connected by direct flights and the number of available seats on any given connection. The combination of the population and mobility layers allows for the subdivision of the world into geo-referenced census areas obtained by a Voronoi tessellation procedure around transportation hubs. These census areas define the subpopulations of the metapopulation modeling structure, identifying 3,362 subpopulations centered on IATA airports in 220 different countries. The model simulates the mobility of individuals between these subpopulations using a stochastic procedure defined by the airline transportation data. Short-range mobility considers commuting patterns between adjacent subpopulations based on data collected and analyzed from more than 30 countries in 5 continents across the world. It is modeled with a time-scale separation approach that defines the effective force of infections in connected subpopulations (see Ref. [18] for details). In other words, short-range mobility is considered at equilibrium in the time scale of long-range patterns. Here, we restricted our analysis to the continental US. To this end, we considered both long and short range mobility patterns limited to the continental US.

- Epidemic Layer defines the disease and population dynamics. The infection dynamics takes place within each subpopulation and assumes a compartmentalization that can be defined according to the infectious disease under study and the intervention measures being considered. As done for the other simulations we considered a SEIR model.

We applied the travel restrictions by multiplying the mobility flows by $p$. However, considering that by construction short-range mobility is encoded in the effective force of infection (in other words in the simulations individuals do not "move" due to short-mobility) we estimate the value of $A(t)$ as:

$$A(t) = p \sum_i \frac{[N_i(t) - I_i(t)]}{N_i(t)} \qquad [9]$$

1. Anderson RM, May RM, Anderson B (1992) *Infectious diseases of humans: dynamics and control.* (Wiley Online Library) Vol. 28.
2. Colizza V, Vespignani A (2008) Epidemic modeling in metapopulation systems with heterogeneous coupling pattern: Theory and simulations. *Journal of theoretical biology* 251(3):450–467.
3. Colizza V, Pastor-Satorras R, Vespignani A (2007) Reaction–diffusion processes and metapopulation models in heterogeneous networks. *Nature Physics* 3(4):276–282.
4. Linkov I et al. (2013) Measurable resilience for actionable policy. *Environmental science & technology* 47(18):10108–10110.
5. Sheffi Y, et al. (2005) The resilient enterprise: overcoming vulnerability for competitive advantage. *MIT Press Books* 1.
6. Cimellaro GP, Reinhorn AM, Bruneau M (2010) Framework for analytical quantification of disaster resilience. *Engineering Structures* 32(11):3639–3649.
7. Cutter SL et al. (2013) Disaster resilience: A national imperative. *Environment: Science and Policy for Sustainable Development* 55(2):25–29.
8. Linkov I et al. (2014) Changing the resilience paradigm. *Nature Climate Change* 4(6):407–409.
9. Vugrin ED, Warren DE, Ehlen MA, Camphouse RC (2010) A framework for assessing the resilience of infrastructure and economic systems in *Sustainable and resilient critical infrastructure systems*. (Springer), pp. 77–116.
10. Baroud H, Ramirez-Marquez JE, Barker K, Rocco CM (2014) Stochastic measures of network resilience: Applications to waterway commodity flows. *Risk Analysis* 34(7):1317–1335.
11. Adger WN, Hughes TP, Folke C, Carpenter SR, Rockström J (2005) Social-ecological resilience to coastal disasters. *Science* 309(5737):1036–1039.
12. Alderson DL, Brown GG, Carlyle WM (2014) Assessing and improving operational resilience of critical infrastructures and other systems. *Stat* 745:70.
13. Barrett CB, Constas MA (2014) Toward a theory of resilience for international development applications. *Proceedings of the National Academy of Sciences* 111(40):14625–14630.
14. Ganin AA et al. (2016) Operational resilience: concepts, design and analysis. *Scientific reports* 6.
15. Perra N, Balcan D, Gonçalves B, Vespignani A (2011) Towards a characterization of behavior-disease models. *PloS one* 6(8):e23084.
16. Hatchett RJ, Mecher CE, Lipsitch M (2007) Public health interventions and epidemic intensity during the 1918 influenza pandemic. *Proceedings of the National Academy of Sciences* 104(18):7582–7587.
17. Markel H et al. (2007) Nonpharmaceutical interventions implemented by us cities during the 1918-1919 influenza pandemic. *Jama* 298(6):644–654.
18. Balcan D et al. (2009) Seasonal transmission potential and activity peaks of the new influenza a (h1n1): a monte carlo likelihood analysis based on human mobility. *BMC medicine* 7(1):1.

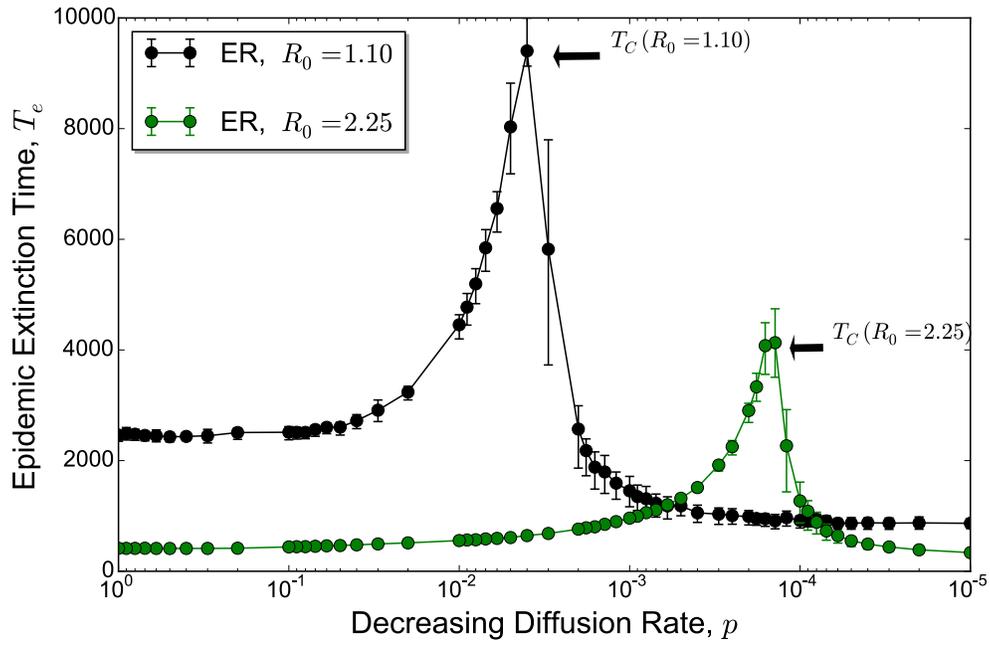

**Fig. S1.** Control time definition. Median value of the epidemic extinction time $T_e$ as function of the diffusion rate $p$. The maximum time correspond to the epidemic control time $T_C(R_0)$.

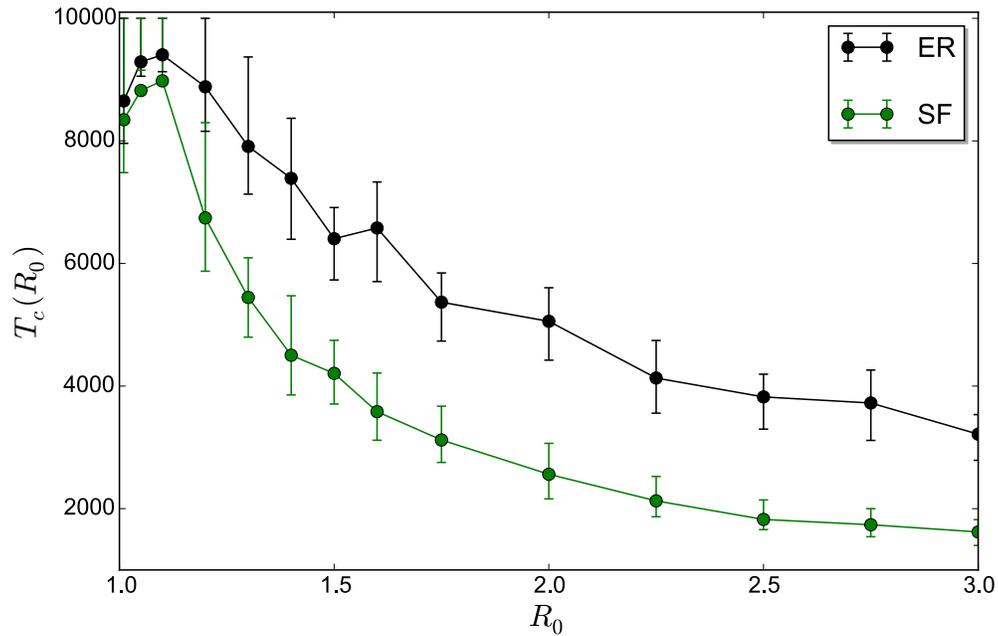

**Fig. S2.** Different values of the control time in both homogeneous and heterogeneous networks for different values of the epidemic reproduction number $R_0$.

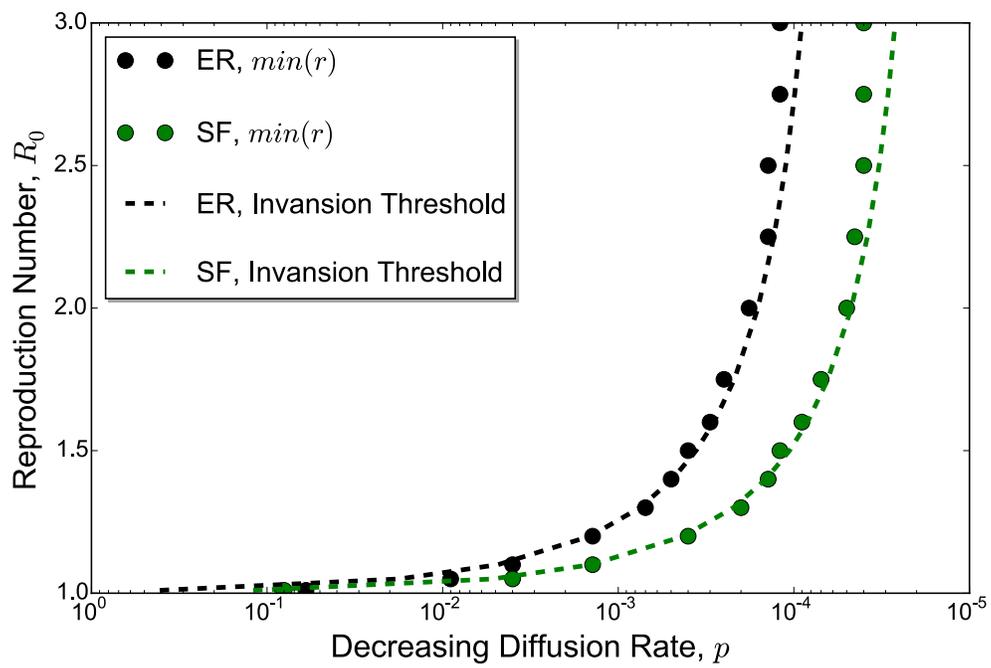

**Fig. S3.** Effect of the network heterogeneity on the system's risk and resilience. The minimum value of the resilience (dots), which corresponds to the theoretical value of the final fraction of diseased subpopulations $D_\infty/V$ at the end of the global epidemic (dotted lines), is shown as a function of the mobility rate p in a homogeneous and heterogeneous networks. The minimum value of the resilience separates the two region of high resilience.

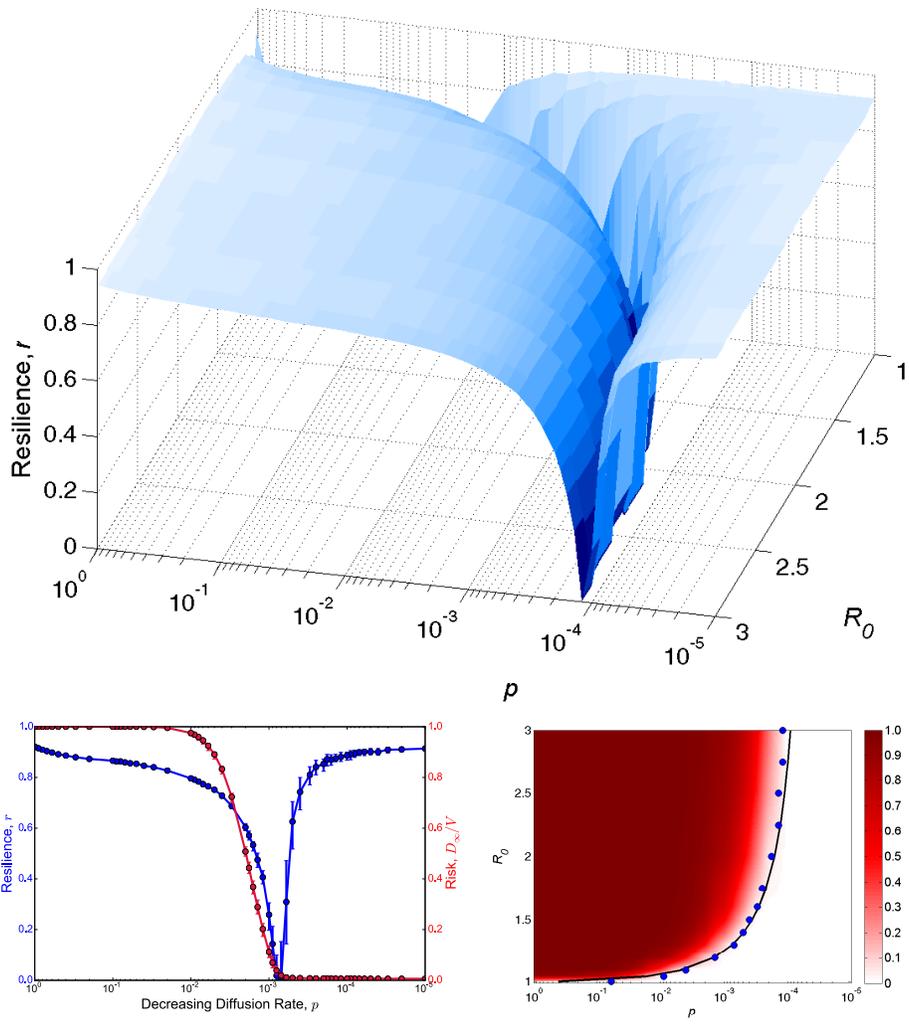

**Fig. S4.** Resilience surface in homogeneous networks in the plane $(p - R_0)$. Figure B refers to $R_0 = 1.3$.

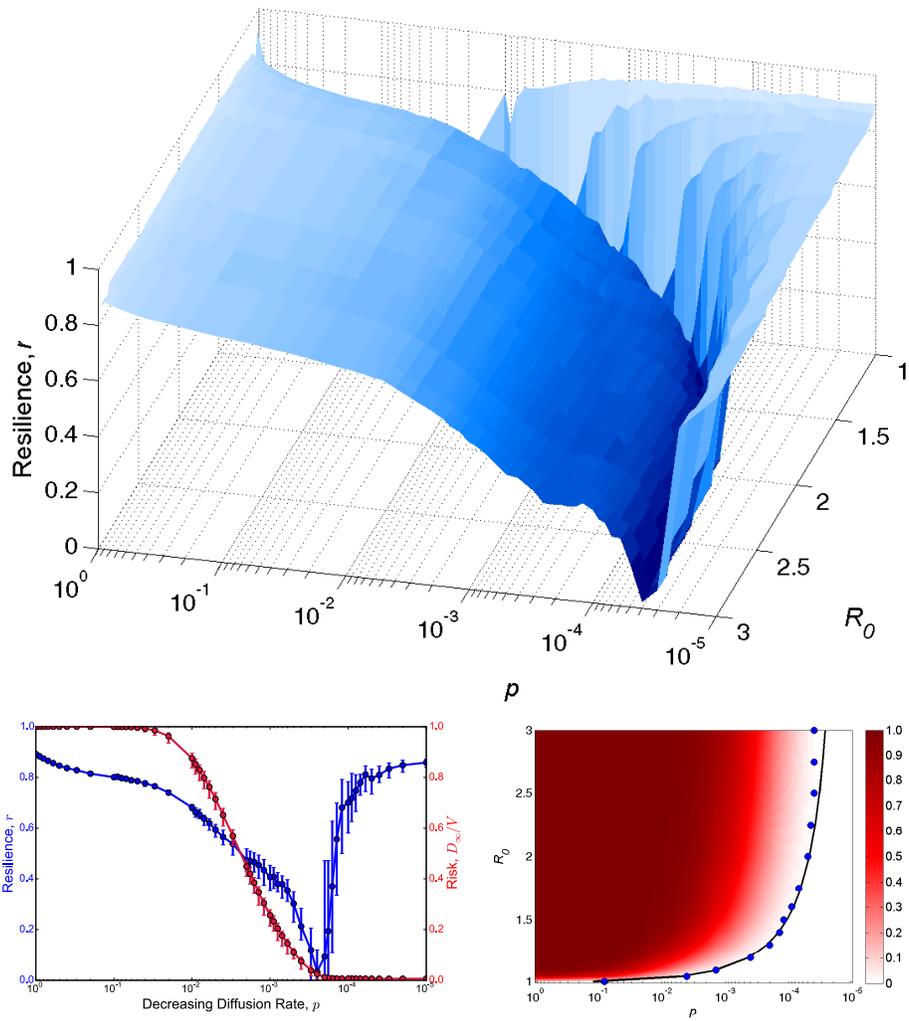

**Fig. S5.** Resilience surface in heterogeneous networks in the plane $(p - R_0)$. Figure B refers to $R_0 = 1.3$.

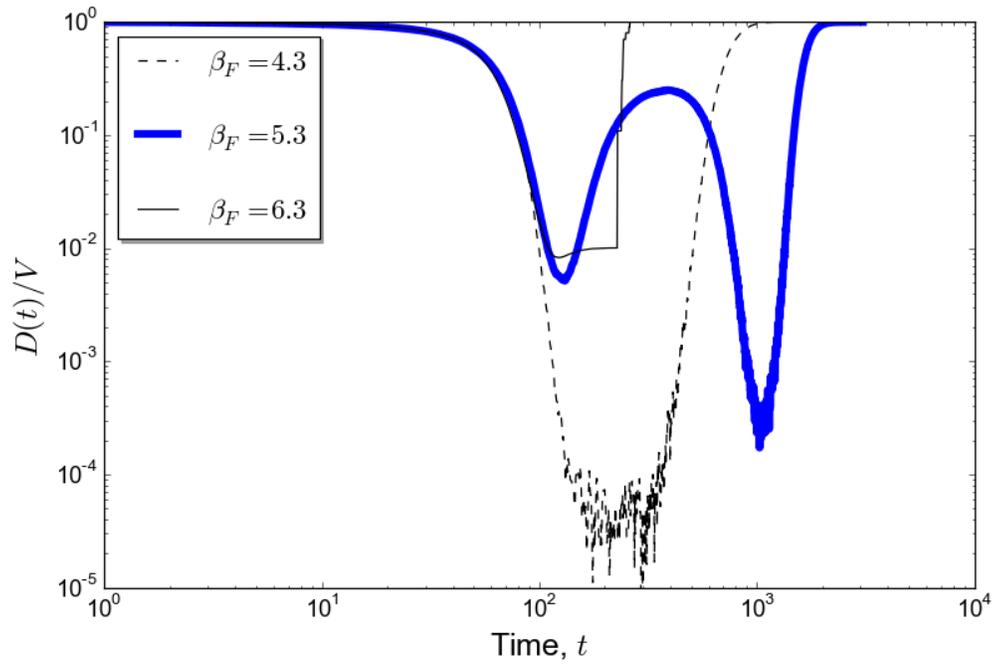

**Fig. S6.** (log-log) Average values of the system's critical functionality for $R_0 = 2$ before (dotted line), over (red line) and after the transition point.

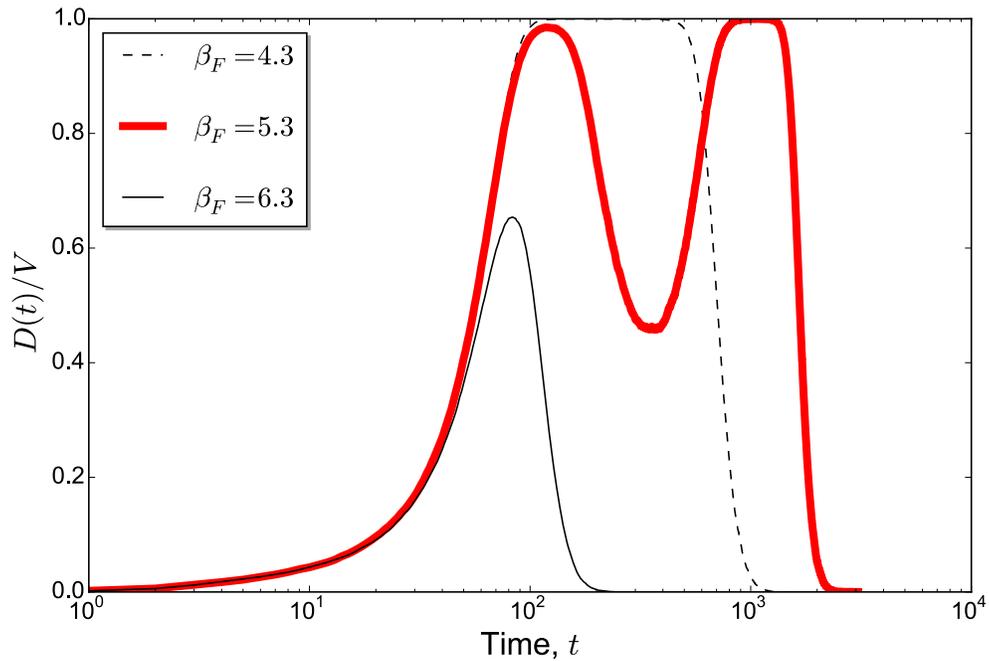

**Fig. S7.** (log-x) Average values of the diseased populations for $R_0 = 2$ before (dotted line), over (red line) and after the transition point.